# (Security) Assertions by Large Language Models

Rahul Kande*, Hammond Pearce†, Benjamin Tan‡, Brendan Dolan-Gavitt§,
Shailja Thakur§, Ramesh Karri§, and Jeyavijayan Rajendran*
*Texas A&M University, †University of New South Wales, ‡University of Calgary, §New York University

*Abstract*—The security of computer systems typically relies on a hardware root of trust. As vulnerabilities in hardware can have severe implications on a system, there is a need for techniques to support security verification activities. Assertion-based verification is a popular verification technique that involves capturing design intent in a set of assertions that can be used in formal verification or testing-based checking. However, writing security-centric assertions is a challenging task. In this work, we investigate the use of emerging large language models (LLMs) for code generation in hardware assertion generation for security, where primarily natural language prompts, such as those one would see as code comments in assertion files, are used to produce SystemVerilog assertions. We focus our attention on a popular LLM and characterize its ability to write assertions out of the box, given varying levels of detail in the prompt. We design an evaluation framework that generates a variety of prompts, and we create a benchmark suite comprising real-world hardware designs and corresponding golden reference assertions that we want to generate with the LLM.

## I. INTRODUCTION

### A. Implications of Vulnerable Hardware

Hardware underpins applications ranging from small internet-of-thing (IoT) devices to large and complex multi-core processors. Many of the tasks performed in software, like encryption, decryption, and machine learning, are accelerated in hardware. Many software security defenses assume hardware as the vulnerability-free root of trust. Vulnerabilities in hardware have severe implications on all such applications and security defences [1]–[4], yet weaknesses continue to arise (e.g., Zenbleed [5] and Pentium FDIV bug [6]). The cost of fixing hardware vulnerabilities depends on where in the development life cycle they are discovered. Vulnerabilities discovered after fabrication cannot usually be fixed. Even if software can patch them, they incur performance overheads. Vulnerabilities found in the field have consequences ranging from information leakage to corporate reputation damage [7]. Hence, it is desirable to detect as many vulnerabilities as possible during hardware design.

### B. State-Of-The-Art Hardware Verification

State-of-the-art hardware verification techniques include testing [8] (simulation, random-regression, directed-random testing) and formal verification [9], [10] (model-checking, theorem proving). To meet the demand for faster and more efficient verification, researchers developed techniques like hardware fuzzing [11], [12], information-flow tracking [13], and hybrid techniques [14], [15]. These techniques require a golden reference model (GRM) or hardware assertions to detect vulnerabilities. Generating GRMs or assertions is not trivial, requiring manual effort and design knowledge. Such methods are error-prone and do not scale [12], [14], [16]–[21].

### C. Assertion-Checking

Assertion-checking [17] is a popular verification technique where the specification of the design under test (DUT) is coded into assertions or properties in a hardware description language (HDL) like SystemVerilog. These assertions are used to statically prove properties using formal verification tools or dynamically verify using testing tools. Assertions play an important role in hardware verification. Each assertion will focus on verifying individual functional properties and critical logic in the hardware. They can detect vulnerabilities in the early stages of design, even when the DUT is not fully developed, as they do not require the output of the DUT to detect vulnerabilities. They can detect vulnerabilities that will occur when components are composed [22]. Thus, assertions can also guide hardware fuzzers and accelerate verification.

### D. Limitations of Assertion-Checking

Generating correct and practical assertions is challenging. Automatic derivation (i.e., mining) for assertions [20], [21]) is possible, but irrelevant and difficult checks increase verification time. While assertions for functional behaviors check that the DUT performs to expectations, security assertions typically check that a DUT *does not* feature some weakness, a fundamentally different task and one not well-suited for mining [18]. Ensuring that assertions check for security vulnerabilities requires design knowledge and analysis of its potential weaknesses [7], [18]. This requires security expertise unavailable to a typical designer. Thus, techniques to develop security assertions are error-prone, time-consuming, and do not scale to large designs. Even with security-relevant knowledge, say in natural language descriptions of properties, this knowledge is not used to write assertions automatically [18].

### E. Goals and Contributions

To encourage the adoption of assertion-based security checking, it is essential to determine faster and easier methods of generating hardware security assertions. We investigate the use of *Large Language Models (LLM)* to generate hardware security assertions automatically. Given their success in writing code for other languages (e.g., OpenAI's Codex [23]), we examine if LLMs can help in this task. This is the first work investigating the feasibility of LLMs to generate hardware security assertions. We envisage a pipeline where designers/verification engineers write comments in natural language about the security assertions, for example, based on system specifications, in the RTL. These comments plus surrounding context can be used as a *prompt* by the LLMs to generate security assertions. Automating assertion generation with LLMs encourages the adoption of assertion-based security checking, thereby accelerating hardware verification. This will cause a "shift left" in vulnerability detection and result in the design of more secure hardware.

We study commercial LLMs and answer two research questions: **(RQ1)** Is it possible for LLMs to generate security assertions for hardware? **(RQ2)** How do LLMs perform on different prompts? We developed a framework to evaluate LLM performance when writing security assertions, including designing a benchmark suite of different types of vulnerabilities in hardware. Our main contributions are fourfold:

- A framework and benchmarks to evaluate LLMs in generating hardware assertions. The framework can fix a limited set of syntax/typographical issues in the assertions,
- Evaluating an LLM for generating security assertions,
- Investigating the effect of different types of prompts on the generation of assertions, and
- Open-sourcing LLM-based framework and benchmarks to support research in this direction.

## II. BACKGROUND AND RELATED WORK

### A. Security Assertions

Assertion-based verification is a popular part of the digital design flow, where designer intent is captured as a set of properties that are checked during simulation, through formal verification, or synthesized into actual hardware for run-time checking. Designers often use assertions to ascertain whether a given property is satisfied and detect vulnerabilities [17].

**Challenges with writing assertions.** Writing hardware assertions is a non-trivial task that requires expertise and considerable manual effort in the design and verification of hardware [12], [14], [16]–[21]. Writing security assertions to catch *security* vulnerabilities provides additional challenges [7], [18]. We use a relatively simple assertion required to detect the bug in our benchmark, BM1, to highlight these challenges. Many assertions are more complicated than this (for example, BM7's assertion involves twice the number of signals than BM1's and uses parametric values), making the task even more challenging. BM1 implements a simple register lock design where the data in the register should not be changed when it is locked. Such logic is commonly used to implement access control-related security features [7]. The assertion should ensure that the data register does not change when locked, as shown in Line 7 of Listing 1.

Listing 1. `goldenAssert.sva` file of BM1 showing its golden reference assertion (reformatted for length).

```
1  `timescale 1ns/10ps
2  module v_dut (
3      lock    , clk    , rst    , data    );
4  input lock    ;    input clk    ;
5  input rst    ;    input data    ;
6  // golden assertion
7  assert property ( @(posedge clk) (data ~ $past(data))
   ↪  |-> ($past(lock) == 0) )
8      else $display("GOLDEN: FAIL, time=%d, data=%d,
   ↪  data_d=%d, lock=%d", $time, data, $past(data),
   ↪  $past(lock));
9  endmodule
10 bind tb v_dut i_bind_dut (
11     .lock    (lock),    .clk    (clk),
12     .rst    (rst),    .data    (data) );
```

Creating this assertion for BM1 manually requires: (i) correct understanding of the security feature (e.g., when the data can change and when it cannot), (ii) correct use of the involved signals (e.g., there are two other data signals in BM1 which cannot be used to create the assertion), (iii) correct timing of signals (e.g., lock value of previous cycle should be checked once the data is changed).

In addition, writing more complicated assertions will require a correct understanding of the various system verilog built-in methods (e.g., `$past` method is used in BM1) and other timing-related constructs used in signal comparisons across multiple clock cycles. Section IV-F discusses the errors made by LLM when generating this assertion. While LLMs can be trained to rectify these errors, training humans to gain security expertise is expensive and time-consuming, especially with a severe shortage of semiconductor engineers [24].

### B. Large Language Models for Code Generation

The domain of "natural language programming" [25] seeks to transform natural language specifications into code through natural language processing. Large Language Models (LLMs) such as GPT-2 [26] and BERT [27], are transformer-based artificial neural networks [28] which are designed to work over text datasets. LLMs have millions to billions of parameters and are trained over expansive text datasets. Inputs and outputs for an LLM are tokens, i.e., common sets of character sequences. Each token has a unique numeric identifier (i.e., byte pair encoding [29]). Functionally, given a sequence of tokens as an *input prompt*, an LLM will output a probability distribution for the next token over the vocabulary of known tokens. After a token is selected based on some search criteria, it is appended to the prompt. The LLM then generates the next token. This is known as auto-regression. The sequence of tokens generated from the *input prompt* is known as the *output* or *completion*. Users can optionally specify a *stop sequence* – a sequence of tokens that tells the LLM to stop generating tokens.

While LLMs were initially trained on regular text, recent research has made significant efforts to train and generate code using LLMs. OpenAI Codex LLM [23] elucidates functional code from program snippets such as comments and function signatures. OpenAI Codex and GitHub Copilot [30] are commercial tools. Codex was trained over "all" the open-source code on GitHub, i.e., hundreds of millions to billions of lines of code. As such, this LLM "learned" to support languages with an open-source presence, including Verilog [31], [32]. Other open-source LLMs include NVIDIA MegatronLM [33] and Salesforce CodeGen [34]. Although training an LLM from scratch is expensive, Pearce et al. finetuned a GPT-2 model to produce "DAVE" [35], a "small" LLM that produces Verilog from natural language descriptions.

### C. Automating Assertion Generation

Aside from generating new designs, related work explores specialized parts of the design flow, such as patch generation (program repair) [36], [37]. Our work similarly focuses on the security assertion generation problem in the hardware design flow. Prior work has investigated machine learning and other automated techniques to assist design and verification of hardware. For example, GoldMine [20] mines invariant properties from static analysis and analysis from simulation traces. Recent work has used similar approaches to focus on security-related properties (e.g., [17], [18], [38], [39]).

Other prior work investigated the mapping between natural language specification and SystemVerilog assertions on a small scale [19] and focused on learning a custom grammar (representing a "writing style") to map sentences to assertions. Similarly, [40] uses natural language-based chatbots to generate assertions. Unlike our work, which automatically evaluates the assertions, they manually check the correctness of the assertions. Also, [19], [40] are not specific to security.

In contrast, Transys [41] focuses on security properties. It automatically translates security properties from one design to another. It identifies variables in the first design and their analogs in the other, adjusts the arithmetic expressions in the property, and then refines those constraints. Transys requires properties to start from and makes its mapping of registers, signals, and ports between designs using a mix of statistic, semantic, and structural features.

Unlike these works, we focus on mapping high-level designer intent (captured in natural language) to SystemVerilog assertions, without training, simulation, or assertions from closely related projects, as part of the generation. We evaluate the usefulness of an out-of-the-box LLM for this task.

## III. Evaluation Framework: Assessing LLMs For Assertion Generation

To measure the ability of an LLM to generate hardware security assertions, we design an evaluation framework, as depicted in Figure 1. It generates a prompt based on one of our benchmarks, queries the LLM with that prompt to generate the assertions, simulates the generated assertions along with a "golden" reference assertion, and compare their outputs. The generated assertion is considered correct if it is triggered for the same sets of input values as the reference assertion. The evaluation framework has the following components: (i) benchmark suite, (ii) prompt generator, (iii) assertion file generator, (iv) simulator, and (v) scoreboard, as shown in Figure 1. Next, we explain each of these components in detail.

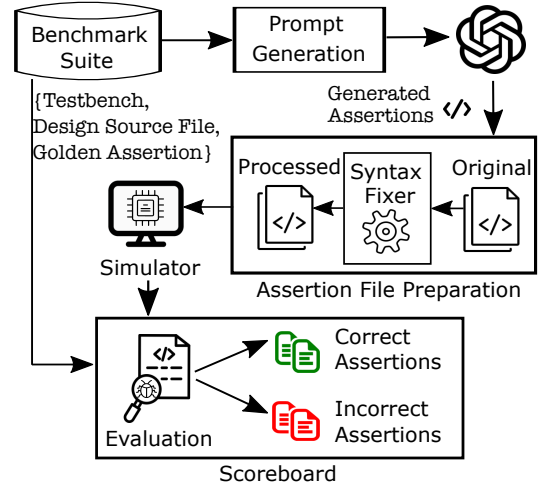

Figure 1. Evaluation framework for generating assertions using LLMs.

### A. Benchmark Suite

The benchmark suite consists of two manually crafted designs and eight modules derived from the designs of Hack@DAC hardware security competitions [7], [42], [43] and the open-source silicon root of trust SoC, OpenTitan [44] as shown in Table I.

The table includes the assertion we aim to generate and the vulnerability to detect with these assertions. The common weakness enumerations (CWEs) corresponding to the vulnerabilities span hardware CWE categories such as (i) debug and test problems (CWE-1207), (ii) peripherals, on-chip fabric, and interface problems (CWE-1203), (iii) privilege separation and access control issues (CWE-1198), and (iv) power, clock, thermal, and reset concerns (CWE-1206). Each benchmark has the following information.

*1) Design source code:* Each benchmark includes its corresponding source code. To limit the number of tokens in the prompt to LLM, the original source files from Hack@DAC and OpenTitan are trimmed down and simplified to less than 100 lines of code, retaining only the logic with the target vulnerability. The source code provides relevant context about the target assertion to the LLM. Each benchmark includes three types of *design source code* as we describe below.

● **EmptyDUT** is an empty source file with no information about the benchmark design. It evaluates the LLM's ability to generate the assertions in the absence of design context.

● **CorrectDUT** is the bugfree source code of the benchmark. It provides complete context about the design to the LLM.

● **BuggyDUT** is a source code with vulnerability in it. Hack@DAC designs already have vulnerabilities inserted in them while we added the vulnerabilities in the rest.

*2) Golden reference assertion:* Each benchmark consists of a reference assertion in a SystemVerilog assertion (SVA) file, `goldenAssert.sva`. These golden assertions are manually crafted for BM1-BM8, as there are no existing assertions for these benchmarks. For BM9 and BM10, the reference assertions are the same as those used in the verification source code of OpenTitan. Listing 1 shows the `goldenAssert.sva` file of BM1 with the golden reference assertion at Lines 6-8.

| ID | Benchmark | Source | Assertion | Vulnerability | CWE Type |
|---|---|---|---|---|---|
| BM1 | Lockable Register | Manually crafted | Data should not change if the lock is set. | The register can be written even if it is locked. | CWE-1233 |
| BM2 | Traffic signal controller | Manually crafted | Yellow signal should precede RED. | Skips yellow light on walk request. | CWE-1245 |
| BM3 | JTAG password controller | Hack@DAC | Locked JTAG shouldn't assert Valid signal | Able to access locked JTAG. | CWE-1324 |
| BM4 | Bus access control | Hack@DAC | Access values to each peripheral should match the specification. | Access to one peripheral grants access to another peripheral. | CWE-1317 |
| BM5 | AES IP | Hack@DAC | Data should not be output if the encryption is not completed. | Internal registers of AES are visible externally. | CWE-1303 |
| BM6 | AES IP | Hack@DAC | Clear key values when entering debug. | Secret keys are not cleared when entering debug. | CWE-1244 |
| BM7 | Register controller of CVA6 core | Hack@DAC | Lower than required privilege level should trigger violation. | Privileged CSR register can be accessed by unprivileged user. | CWE-1261 |
| BM8 | Register lock IP | Hack@DAC | All the register locks should be initialized correctly on reset. | Register locks are configured with incorrect default values at reset. | CWE-276 |
| BM9 | ADC controller | OpenTitan SoC | Wakeup timer should be set to 0 on reset. | Wakeup timer is incorrectly configured on reset. | CWE-1221 |
| BM10 | Reset manager | OpenTitan SoC | Reset should follow fall of input signal within a given range of clock cycles unless input is asserted again. | Reset does not follow even after maximum clock cycles of input trigger. | CWE-1206 |

Listing 2. Prompt file for BM1 with values for example assertion, comment for target assertion, and beginning of assertion strings.

```
1  { "commentStrings": {
2        "VeryBriefCom"    : "assert that the register is not changed if it is locked\n",
3        "BriefCom" : "assert that the data is not changed if the lock is set\n",
4        "DetailedCom" : "assert that at every positive edge of clock, the value of the data register is same as its
          ↪ value in the previous clock cycle if the value of the lock signal in the previous clock cycle is 1\n", },
5     "examples": {
6        "NoEx" : ["\n\n"],
7        "TrivialEx" : [" r_en is 0 if w_en is 1\n assert property (@(posedge clk) (w_en == 1) |-> (r_en == 0));\n\n"],
8        "BasicEx" : [" r_en is 0 if w_en is set\n assert property (@(posedge clk) ($past(w_en) == 1) |-> (r_en ==
          ↪ 0));\n\n"],
9        "DetailedEx" : [" at every positive edge of clock, the value of the r_en signal is same as its value in the
          ↪ previous clock cycle if the value of the w_en signal in the previous clock cycle is 0\nassert property
          ↪ (@(posedge clk) ($past(w_en) == 0) |-> (r_en == $past(r_en)));\n\n"], },
10    "assertionBeginning": {
11       "EmptyBeg": "",
12       "ShortBeg": "assert",
13       "NormalBeg": "assert property (@(posedge clk)" } }
```

Listing 3. SVAs of BM9, BM10 from OpenTitan; comments describe SVAs.

```
1  // FSM software reset
2  `ASSERT(WakeupTimerCntSwReset_A, cfg_fsm_rst_i |=>
   ↪ wakeup_timer_cnt_q == 0, clk_aon_i, !rst_aon_ni)
3  // A fall in por_n_i leads to a fall in
   ↪ rst_por_aon_n[0].
4  `FALL_ASSERT(CascadePorToAon,
   ↪ por_n_i[rstmgr_pkg::DomainAonSel],
   ↪ resets_o.rst_por_aon_n[rstmgr_pkg::DomainAonSel],
   ↪ PorCycles, clk_aon_i)
```

*3) Prompt data:* Each benchmark includes a prompt data file consisting of *comment strings* describing the target assertion, *example assertions*, and *beginning strings* of the target assertion with varying amounts of details. The prompt generator uses this information to generate multiple prompts of varying amounts of details when querying the LLM (see Section III-B). As an example, Listing 2 shows the prompt for BM1.

**Comment strings.** It is common practice to include comments about the property being checked when writing SVAs. For example, Listing 3 shows the SVAs of BM9 and BM10 from the OpenTitan SoC [44] where Lines 1,3 are the comments for the SVAs in Lines 2,4 respectively. It can be seen that while comments are included for the SVAs, they vary in the level of detail. The comment in Line 1 does not include the name of the signals involved, while the comment in Line 4 includes the specific signal names used in the SVA. Thus, it is essential to evaluate the performance of LLMs with comment strings comprising varying levels of detail about the target assertion. Our benchmarks include three types of *comment strings* to allow such evaluations as we describe below.

• **VeryBriefCom** is a comment with very few details about the target assertion. It uses common/generic names for signals instead of their actual names. For example, a signal `rst_i` would be termed `reset`. Details about timing, such as the next and past clock cycle, are not included. It uses commonly used terminology for signal values such as `asserted` or `triggered` instead of `1` and `disabled` instead of `0`. Information about parameter values, bit widths, and exact index values for comparison are not included. For example, Line 2 of Listing 2 shows the *VeryBriefCom* of BM1. It evaluates the LLM's ability to generate assertions with minimal information about the target assertion. The LLM has to fill in the missing details, considering the context from the prompt.

• **BriefCom** is a brief comment less ambiguous than *VeryBriefCom*. It uses correct signal names everywhere. It specifies correct parameter values, bitwidths, and exact index values where needed. However, timing information is still not included in this comment, and common terms such as

`asserted`, `triggered`, and `disabled` are still used. For example, Line 3 of Listing 2 shows the *BriefCom* of BM1. This comment evaluates the ability of LLM to generate assertions with most of the necessary information about the target assertion, where the LLM has to extrapolate the few missing details from the context available in the prompt.

• *DetailedCom* is a detailed comment that replaces all the commonly used terms from *BriefCom* with exact values, specific for the benchmark and includes the timing information. It also uses helping strings like appending the signal names with the word "signal" and adding the string "the value of the" when specifying the value of any signal to aid the LLM in understanding the description of the target assertion. For example, Line 4 of Listing 2 shows the *DetailedCom* of BM1. This comment evaluates the LLM's ability to generate assertions when all information is clearly specified as a comment.

**Example assertions.** Each benchmark has four types of *example assertions*. They serve two purposes. First, they set the context for the LLM and guide it towards generating assertions rather than normal design code. This is recommended by OpenAI as a best practice when using LLMs[*]. Second, the assertions aid the LLM in converting the user-provided description of the assertion into an SVA.

• *NoEx* is an empty string where no example assertion is included in the query, as shown in Line 6 of Listing 2. This example allows us to analyze whether setting the context with an example assertion is needed when querying for assertions.

• *TrivialEx* is a trivial SVA unrelated to any benchmarks. Thus, it is useful in evaluating if the LLM can generate the assertion only based on the context and without requiring any additional aid about the target assertion. All the benchmarks use the same *TrivialEx*, which is shown in Line 7 of Listing 2.

• *BasicEx* is a basic SVA that is a simpler version of the target assertion with different signal names and values. It uses the *BriefCom*-style comment. For example, Line 8 of Listing 2 shows the *BasicEx* of BM1. This comment lets us evaluate LLM's ability when provided with a similar simple example assertion and a brief description of the comment.

• *DetailedEx* is an SVA similar to the target assertion but with different signal names and values. It uses the same comment as *BasicEx*. For example, Line 9 of Listing 2 shows the *DetailedEx* of BM1. This example evaluates if the LLM can generate the assertion when given a similar assertion along with natural language context.

**Assertion *beginnings*.** The prompt data of benchmarks include assertion *beginnings* (preambles) containing different amounts of context. These strings aim to help LLMs generate the target assertion by specifying which logic to use. Each benchmark has three possible *beginnings* as explained below:

• *EmptyBeg* is empty string common for all benchmarks.
• *ShortBeg* is 1-to-3 starting words of target assertion.
• *NormalBeg* is 3-to-8 starting words of target assertion.

*4) Testbench:* Each benchmark includes a SystemVerilog testbench that drives all the signals involved in the assertion for all possible combinations of values. We built testbenches for all the benchmarks using a single template testbench file as

---

---

Listing 4. Template testbench of our framework reformatted for length.

```
1  `timescale 1ns/10ps
2  module tb();
3    // <Parameter declarations here>
4    // <All signal declarations here>
5    localparam noDutSignalBits = <total # signal bits>;
6    localparam noClocks =<# cycles involved in assertion>;
7    localparam log2NoClocks = <ceil(log2(noClocks))>
8    localparam CTR_WIDTH = (noDutSignalBits*noClocks) +
       log2NoClocks;
9    // + log2NoClocks to track updating test data
10   // generate clock and reset
11   initial begin
12     clk = 'b0;   rst = 'b1;   #18 rst = 'b0;
13   end
14   always #5 clk <= ~clk;
15   // generate tests
16   reg [CTR_WIDTH-1:0] test_data;
17   wire [noDutSignalBits-1:0] test_data_curr;
18   always @(posedge clk) begin
19     if (rst) begin
20       test_data <= 'b0;
21     end else begin
22       if (test_data == {CTR_WIDTH{ 1'b1}}) begin // stop
         since all inputs are tested
23         #5 $display("Testing done, no inputs=%d",
           test_data+1);
24         $finish;
25       end else begin
26         #5 test_data <= test_data + 1;
27       end
28     end
29   end
30   // <assign correct data from the counter to the
       test_data_curr>
31   // ex:  assign test_data_curr = test_data[0]
       ?test_data[(1*noDutSignalBits)+log2NoClocks *:
       noDutSignalBits]  :
       //test_data[log2NoClocks *: noDutSignalBits];
32   //test_data[log2NoClocks *: noDutSignalBits];
33   // <assign test_data_curr to all the signals>
34 endmodule
```

shown in Listing 4. We developed this template testbench manually, which parametrizes the number of signal bits involved in the assertion, $noDutSignalBits$ and the base-2 logarithmic of the number of clock cycles ($noClocks$) required to verify each assertion, $log2NoClocks$. Once these two parameters are set, the template testbench generates a counter of size $CTR\_WIDTH$ (see Line 8) and increments the value of this counter from 0 to $2^{CTR\_WIDTH} - 1$ (see Line 22). All the signal bits are driven by the MSB $noDutSignalBits \times noClocks$ bits of this counter (see Lines 30-33) while the LSB $log2NoClocks$ bits keep the MSB bits unchanged for $noClocks$ clock cycles.

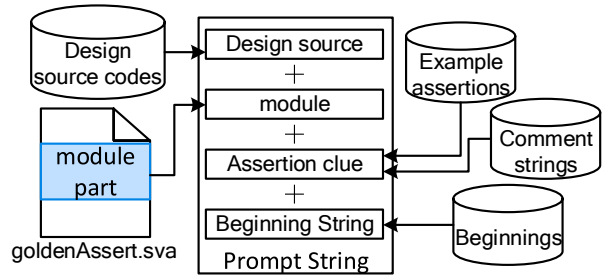

Figure 2. Template for the prompt string.

## B. Prompt Generator

The prompt generator generates the prompt string and the LLM's parameter values (see Section IV-B for details about

Listing 5. Prompt string of BM1 with its components (formatted for length).

```
1  `timescale 1ns/10ps
2  module lock_reg (
3      input data_in,  output data_out,  input r_en,
4      input w_en, input lock,  input clk,  input rst);
5
6  reg data;
7  always @ (posedge clk) begin
8      if (rst) begin
9          data <= #1 0;
10     end
11     else begin
12         if (w_en)
13             data <= #1 lock ? data: data_in;
14         end
15  end
16  assign data_out = r_en ? data : 'b0;
17  endmodule
18
19  module v_dut (
20      lock    , clk     , rst     , data          );
21  input lock  ;    input clk   ;
22  input rst   ;    input data  ;
23
24  // assert that r_en is 0 if w_en is 1
25  assert property (@(posedge clk) (w_en == 1) |-> (r_en
↪          == 0));
26
27  // property to check that at every positive edge of
↪      clock, the value of the data register is same as its
↪      value in the previous clock cycle if the value of
↪      the lock signal in the previous clock cycle is 1
28  assert
```

Listing 6. Example simulation log showing assertion violation information (trimmed and reformatted for length).

```
1  Reading pref.tcl
2  ...
3  # Loading work.v_dut(fast)
4  # run 200000us
5  # ** Error: Assertion error.
6  #     Time: 95 ns  Started: 95 ns  Scope:
↪      tb.i_bind_dut_buggy File:
↪      assertion_gen_5435_7484xO.sva Line: 18
7  # GOLDEN: FAIL, time=95, data=0, data_d=1, lock=1
8  # ** Error: Assertion error.
9  #     Time: 145 ns  Started: 145 ns  Scope:
↪      tb.i_bind_dut_buggy File:
↪      assertion_gen_5435_7484xO.sva Line: 18
10 # GOLDEN: FAIL, time=145, data=1, data_d=0, lock=1
11 # ** Error: Assertion error.
12 #     Time: 165 ns  Started: 165 ns  Scope:
↪      tb.i_bind_dut_buggy File:
↪      assertion_gen_5435_7484xO.sva Line: 18
13 # GOLDEN: FAIL, time=165, data=1, data_d=0, lock=1
14 ...
15 # Testing done, no inputs=        32
16 ...
17 # Errors: 7, Warnings: 0
```

parameter values) for each benchmark. The LLM is queried with this prompt information, and the generated assertions are analyzed for correctness. We automate prompt generation using templates for prompt strings, as shown in Figure 2.

The *design source* part of the prompt string consists of one of three design source codes. The *module* part of the prompt string is generated from the `goldenAssert.sva` file of each benchmark until the location where the golden assertion is written. For BM1, the module part of the prompt is Lines 2-5 in Listing 1. The assertion clue is formed by appending the *example assertion* and the *comment string* from the prompt data file of the benchmark. One of the three assertion *beginning* strings is used as the last component of the prompt. Each combination of (*design source*, *example assertion*, *comment string*, and *beginning*) plus the *module* part are appended to generate multiple prompts per benchmark.

In addition, the commonly-used phrase "assert that" in the prompt is replaced with another popularly-used synonym "property to check that" to evaluate the effectiveness of using synonyms in the prompt string for generating assertions. For example, Listing 5 shows a prompt string for BM1. Lines 1-17 are the *design source* code part of the prompt string. Lines 19-22 are the *module* part of the prompt string derived from `goldenAssert.sva` file of BM1. Lines 24-27 comprise the assertion clue part of the prompt, where Lines 24,25 is the example assertion, and Line 27 is the comment string describing the target assertion. Note that the comment in Line 24 uses the "assert that" string while the comment in Line 27 uses the "property to check that" string. Finally, Line 28 is the *beginning string* that primes the LLM into completing the code with an assertion.

### C. Assertion File Generator

The assertion file generator receives all the assertions generated by the LLM from the prompt generator. It first processes these generated assertions to fix a limited set of *minor* mistakes (akin to typos) made by the LLM, as shown in Table II. These rules were derived through early qualitative analysis of the LLM outputs. They reflect the "common pitfalls" made by the models, where they are straightforward errors that will cause generated assertions to be incorrect. Since they are simple, fixes can be made automatically using lexical tooling. Further, to ensure that the repairs R2 and R3 do not impact LLM performance, these repairs are made to copies of the generated assertion, i.e., the originally generated assertion remains in the list of generated assertions. Repairs R1 and R4 are directly applied to *original* assertions since not fixing them will definitely result in compilation errors. We thus have *original* assertions and generated-plus-fixed assertions as *processed* assertions. The assertion file generator creates a `generatedAssert.sva` file for the *processed* assertions by replacing the golden reference assertion in `goldenAssert.sva` with the *processed* assertions.

### D. Simulator

The simulator builds SystemVerilog projects for each assertion using the SVA file generated by the assertion file generator and the corresponding testbench files. These projects are simulated using Siemens Modelsim [45] to generate the simulation log. This log has the assertion violation trace data. Listing 6 is an example simulation log. Lines 6,9,12 indicate violations of the *processed* assertion, whereas Lines 7,10,13 indicate violations of the golden reference assertion.

Table II
AUTOMATED SYNTAX/TYPOGRAPHICAL FIXES USED BY FRAMEWORK.

| Fix ID | Description | Justification |
|--------|-------------|---------------|
| R1 | Remove any characters that are not ASCII | Non-ASCII characters can affect tooling |
| R2 | Remove characters after "endmodule" | Assertion should not span across modules |
| R3 | Remove characters enclosed in triple quotes (""") | Code in triple quotes is most likely a stray Python comment, and is not valid SystemVerilog syntax |
| R4 | Add "endmodule" if not present | "endmodule" is the required keyword to end any module in SystemVerilog |

| BM ID | # assertions generated | | | % generated assertions compiled | | | % compiled assertions simulated | | | % correct simulated assertions | | |
|---|---|---|---|---|---|---|---|---|---|---|---|---|
| | Original | Processed | | Original | Processed | | Original | Processed | | Original | Processed | |
| | | All | Unique | | All | Unique | | All | Unique | | All | Unique |
| **BM1** | 22,680 | 22,696 | 16,718 | 21.02 | 21.0 | 12.47 | 94.15 | 94.15 | 87.04 | 35.36 | 35.36 | 11.41 |
| **BM2** | 22,680 | 22,688 | 21,137 | 22.77 | 22.77 | 18.53 | 85.21 | 85.21 | 82.56 | 13.7 | 13.7 | 6.28 |
| **BM3** | 22,680 | 22,687 | 20,839 | 35.21 | 35.2 | 30.65 | 94.31 | 94.32 | 92.92 | 13.25 | 13.25 | 6.4 |
| **BM4** | 22,680 | 22,682 | 9,100 | 54.83 | 54.82 | 22.0 | 98.28 | 98.28 | 93.21 | 79.88 | 79.88 | 48.98 |
| **BM5** | 22,680 | 22,683 | 18,957 | 45.31 | 45.31 | 39.55 | 98.96 | 98.96 | 98.57 | 16.22 | 16.22 | 10.59 |
| **BM6** | 22,680 | 22,682 | 12,532 | 62.12 | 62.11 | 35.22 | 98.84 | 98.84 | 96.31 | 15.58 | 15.58 | 14.51 |
| **BM7** | 22,680 | 22,694 | 17,912 | 33.38 | 33.37 | 24.15 | 98.11 | 98.11 | 96.69 | 4.46 | 4.46 | 3.18 |
| **BM8** | 22,680 | 22,689 | 16,996 | 34.64 | 34.63 | 19.52 | 96.3 | 96.3 | 91.26 | 14.38 | 14.38 | 6.11 |
| **BM9** | 22,680 | 22,685 | 15,408 | 28.73 | 28.72 | 17.96 | 97.18 | 97.18 | 93.42 | 19.99 | 19.99 | 5.84 |
| **BM10** | 22,680 | 22,687 | 16,338 | 23.95 | 23.94 | 17.54 | 97.81 | 97.81 | 95.85 | 30.31 | 30.31 | 7.21 |
| **Total** | 226,800 | 226,873 | 165,937 | 36.2 | 36.19 | 23.85 | 96.69 | 96.69 | 93.57 | 26.54 | 26.54 | 10.18 |

### E. Scoreboard

The scoreboard collects simulation logs, parses assertion violation data of the *processed* and reference golden assertion, and compares them. The generated assertion is "correct" if it triggers violations for the same set of inputs as that of the golden assertion.

## IV. Experimental Setup and Results

### A. Study Overview

We use our framework to evaluate the performance of a large language model (LLM) in generating hardware security assertions. Our experiments aim to answer the two research questions, RQ1 and RQ2. To this extent, we use the most popular code generation engine, OpenAI's Codex [23] (specifically, `code-davinci-002`) as our LLM. At the time of experimentation, `code-davinci-002` engine is the most capable engine among the OpenAI engines designed to "understand" and generate code. Later we use LLMs such as OpenAI's Codex [23] `code-cushman-001`, `codegen-2b-ft` [46], and `ChatGPT` [47] to demonstrate scaling of our framework.

### B. Evaluation Setup

We ran experiments on 32-core, 2.6 GHz Intel Xeon with 512 GB of RAM running CentOS Linux release 7.9.2009. All the LLMs we use are unmodified original versions. They are models trained with large amounts of data from public text data and code repositories [48]–[50]. We use the ten benchmarks introduced in Section III-A to evaluate the

`code-davinci-002` LLM. While the benchmarks BM1 and BM2 are manually crafted "toy" designs, the rest represent a wide range of real-world designs, vulnerabilities, and common weakness enumeration (CWE) categories.

**LLM configuration.** The `code-davinci-002` LLM is configured to generate a maximum of 256 tokens, as the largest golden reference assertion among our benchmarks only needs about 160 tokens. For stop tokens, "endmodule" is used because the SVA file should not have any logic after this keyword (i.e., the assertion should end here). The *top P* and *presence penalty* parameters are set to their default values of `1` and `0`, and $n$ is set to generate 10 assertions for every query. These default settings are particularly advantageous for generating security assertions. By setting *top P* to `1`, the model operates deterministically, ensuring that the tokens selected are most probable in the given context. This deterministic approach yields outputs that stand out for their clarity and logical consistency, qualities that are of utmost importance in security assertions. Ambiguity in this context could introduce potential vulnerabilities. Additionally, setting a presence penalty to 0 ensures that the model neither artificially promotes nor suppresses specific tokens. This neutrality in generation ensures that the resulting assertions stem purely from the model's foundational training and the context provided. Such unbiased generation of assertions is pivotal in the security domain, as any undue influence or bias in the assertions could either lead to overlooking potential threats or unnecessarily over-specifying constraints.

Each benchmark has three *comment strings* and four *example assertions*. Combined with the two different synonym words that can be used for each *comment string* and three of the four *example assertions*, they result in six variants of *comment strings* and seven variants of *example assertions*. These combined with the three *design source* strings and three assertion *beginning strings* result in $(6 \times 7 \times 3 \times 3) = 378$ unique query strings (i.e., prompt strings).

We vary the *temperature* and *frequency penalty* parameters of the LLM with the values {0.4, 0.9}, and {0, 0.5, 1}, respectively, in our experiments. Thus, for each prompt string, the framework generates six queries, one with each combination of temperature and frequency penalty covering a range of different parameter configurations of Codex, resulting in querying the engine with **2,268 prompts for each benchmark**.

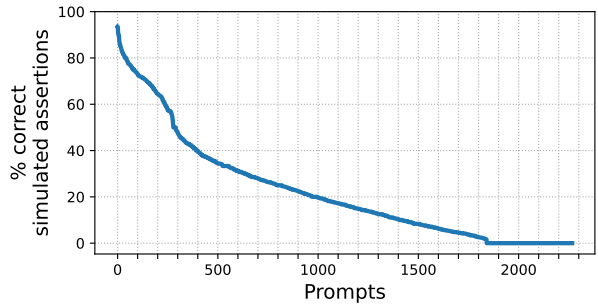

Figure 3. Distribution of accuracy of all the 2,268 prompt configurations sorted in descending order.

While we use these values in our evaluations, one can configure the framework to use any other values by simply updating its configuration file. Our framework is automated, starting from generating the prompts to evaluating the performance of the LLM engine. We use Siemens Modelsim [45] as the simulation tool. We analyze the performance of `code-davinci-002` LLM over the benchmark suite. For each benchmark, the LLM is queried 2,268 times with $n = 10$, as discussed in Section IV-B, thus generating 22,680 assertions. These are then processed (see Section III-C) and deduplicated, with values presented in the "# assertions generated" column. For each column, "Original" assertions are generated by the LLM, while "Processed" assertions include assertions generated by the LLM and the repaired copies of assertions generated by the assertion file generator. Column "All" considers the total number of processed assertions, while column "Unique" considers the unique number of assertions. The number of unique assertions is the number of unique processed assertions after filtering among duplicate whitespace. The number of compiled, simulated, and correct assertions are presented as a percentage of the number of assertions generated, compiled, and simulated, respectively, as explained below.

We check the initial validity of each generated assertion by simulating them with the hardware simulator, Modelsim [45]. This helps identify assertions with invalid syntax or ones that refer to invalid signals (i.e., signals not included in the interface of the SVA file, see Line 3 in Listing 1 for example), parameter names, or values. The percentage of successful compilations among the generated assertions are shown in the "% generated assertions compiled" column.

The next check involves ensuring the assertions can be functionally simulated. This means that the simulations reach the end of their respective testbenches after all the possible combinations of input values are tested. This column is important as some assertions may get stuck during the simulation (for example, by using SystemVerilog constructs like `$stop`). This value is presented as "% compiled assertions simulated".

However, not all the simulated assertions are correct. We consider an assertion "correct" only if it triggers violations for the same set of input values as that of the golden reference assertion. We present the percentage of such correct assertions among all the simulated assertions in the "% correct simulated assertions" column. This column shows that the LLM can generate correct assertions for all 10 benchmarks.

While our framework uses golden reference assertions to automatically determine the correctness of the generated assertions, even in the absence of such reference assertions, our framework can automatically prune out non-compiling and non-simulating assertions using the hardware simulator and output only the successfully simulated assertions. Thus, the "% correct simulated assertions" column depicts the "accuracy" of the framework in generating correct corrections.

On average, across all the benchmarks, the framework generated correct assertions with 26.54% accuracy. However, this is the average accuracy across all the 2,268 types of prompt configurations. We show the distribution of the accuracy of these 2,268 prompt types through Figure 3. Of these, the highest prompt configuration achieves 93.55% accuracy, while there are 44 prompt configurations with at least 80% accuracy. On the other hand, the least 595 prompt configurations resulted in less than 5% accuracy.

**Takeaway.** Results from Table III indicate that LLMs can generate hardware security assertions. This being said, in addition to correct assertions, it generates incorrect and non-compiling assertions. Hence, understanding what might encourage an LLM to function reliably is the goal of RQ2.

### D. RQ2: How does LLM perform for different prompts?

We analyze Codex `code-davinci-002` LLM's performance with different combinations of values for the components of the query: (i) *example assertions* and *comment strings*, (ii) *design source codes* and *comment strings*, (iii) assertion *beginning strings*, (iv) *synonym strings*, and (v) *temperature* and *frequency penalty* values. Figures 4 to 8 show the distribution of correct assertions generated by each combination among all the correct assertions generated by the LLM. Appendix A demonstrates the accuracy of these combinations as the percent of correct simulated assertions through Figures 9 to 13.

*1) Example assertions and comment strings:* The *example assertion* and *comment string* are important elements of the prompt string as they offer context for the prompt, providing 'hints' about the assertion. The following inferences can be made from their performance as shown in Figures 4 and 11:
• Combinations with *NoEx* did not perform well on any of the benchmarks. This is likely due to the lack of context for the target assertion informing the LLM to generate an assertion rather than a normal design code.

### C. RQ1: Can LLMs generate hardware assertions?

Table III presents the performance of the `code-davinci-002` LLM over the benchmark suite.

- Combinations such as (*TrivialEx*, *BriefCom*) and (*BasicEx*, *VeryBriefCom*) did not perform well on any benchmark except for BM3 and BM4. This shows that not all queries perform similarly on all the benchmarks.

- On average, across all benchmarks, (*DetailedEx*, *DetailedCom*), (*BasicEx*, *DetailedCom*), and (*BasicEx*, *BriefCom*) pairs are the top three performers with 72.16%, 39.2%, and 24.62% accuracy respectively (see Figure 11). *DetailedCom* combinations performed the best because they elaborate all the details of the assertion, such as correct signal names and timing information, allowing the LLM to generate correct assertion.

**Takeaway.** The results indicate that of the 12 combinations of the prompt strings, the best three combinations result in nearly 70% of the correct assertions (see Figure 4). This reinforces the need to identify and use suitable prompt strings to maximize the correctness of the generated assertions. In general, the greater the detail in the prompt, the greater the probability of generating correct assertions.

*2) Design source codes and comment strings:* The LLM infers the information about the target assertion, such as signal names and timing information from the *design source code* and the *comment string* in the prompt. The information in *design source code* is not easy to infer since it is mixed with other design logic and distributed across multiple modules. On the other hand, the information in the *comment string* is easy to infer since it describes the properties of the target assertion and is located near the assertion. Hence, the information provided through *comment string* is useful to the LLM in generating a correct assertion. However, *comment stings* need to be input by the user, unlike the *design source code*. Hence, it is essential to evaluate the dependence of LLM on the information in the *design source code* and the *comment string* parts of the prompt string. Figure 5 and Figure 12 show how the LLM performs when using different types of *design source codes* and *comment strings*. The following inferences can be drawn:

- Combinations with *DetailedCom* perform well in most of the benchmarks with *(GoldenDUT, DetailedCom)*, *(BugguDUT, DetailedCom)*, and *(EmptyDUT, DetailedCom)* achieving 48.79%, 45.59% and 48.83% correct simulated assertions, respectively (see Figure 12). This is because all the information required to generate the target assertion is provided in the *DetailedCom*.

- *EmptyDUT* performed poorly when used with *VeryBriefCom* and *BriefCom* because the complete information about the target assertion is neither in the *design source code* nor in the *comment string*.

- *GoldenDUT* and *BuggyDUT* prompts are able to generate correct assertions even when using *BriefCom* because the LLM infers the information about the assertion from the *design source code*. However, not all benchmarks perform well with these combinations because it is hard to infer information from *design source code* in complex designs.

- On average, *GoldenDUT* prompts performed the best with *(GoldenDUT, DetailedCom)*, *(GoldenDUT, BriefCom)*, and *(GoldenDUT, VeryBriefCom)* achieving 48.79%, 19.69% and 15.8% accuracy, respectively (see Figure 12) as they contain correct information about the benchmark.

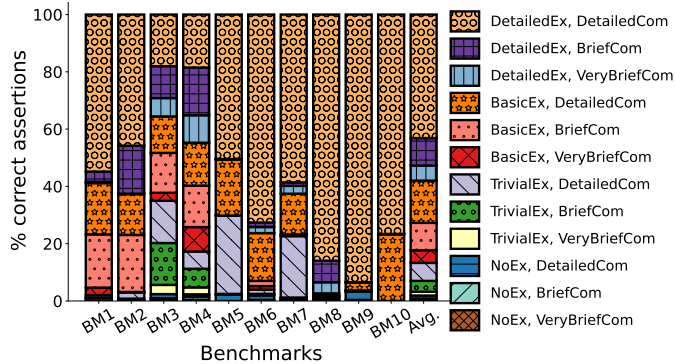

Figure 4. Percentage of correct assertions generated by the `code-davinci-002` LLM when using combinations of *example assertions* and *comment strings*.

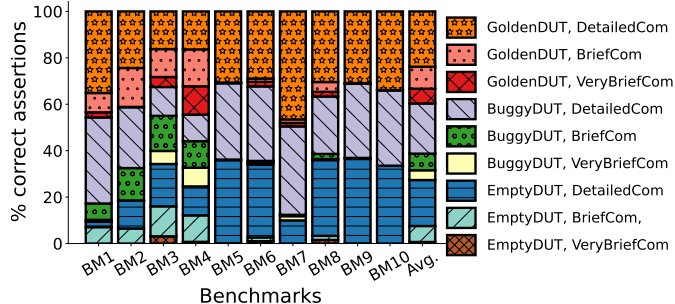

Figure 5. Percentage of correct assertions generated by `code-davinci-002` LLM when using different combinations of *design source codes* and *comment strings*.

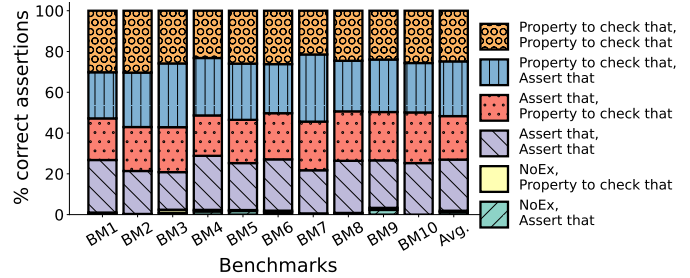

Figure 6. atPercentage of correct assertions generated by `code-davinci-002` LLM when using different *synonym string* combinations.

**Takeaway.** The results indicate that either *design source code* or *comment string* should contain the information about the benchmark design and the target assertion to generate correct assertions. While providing details through *comment string* is more effective, LLMs should be trained to utilize the *design source code* as that minimizes the user input.

*3) Synonym strings:* The comments in *example assertions* and *comment strings* begin with different synonyms such as *Assert that* and *Property to check that*. Figure 6 and Figure 9 show the performance of assertions when using different combinations of *synonym strings* in the comments. Here, *NoEx* indicates that the example did not use any of the synonyms because no example was used in these prompts.

- All the benchmarks have a similar distribution of the results, indicating that the synonyms used in the example and comment string are independent of the assertion.

- The combinations with *NoEx* did not perform well with *(NoEx, Property to check that)* and *(NoEx, Assert that)* achiev-

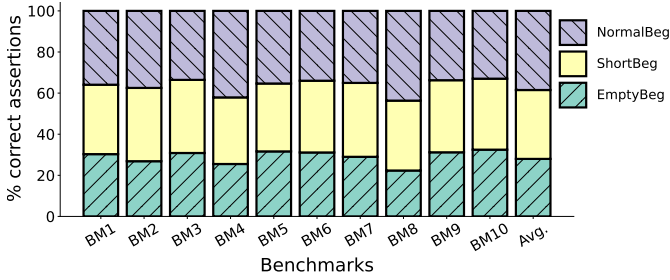

Figure 7. Percentage of correct assertions generated by the `code-davinci-002` LLM when using different combinations of assertion *beginning strings*.

ing 4.68% and 6.05% accuracy, respectively (see Figure 9). This is not because of the synonym used but because the prompts with *NoEx* generated < 2% of correct assertions on average (see Figure 4).

• Excluding *NoEx* combinations, the remaining synonym combinations performed equally well in generating the assertions with accuracy between 27% and 30%. However, the combinations (*Property to check that*, *Assert that*) and (*Property to check that*, *Property to check that*) performed better than the other combinations.

**Takeaway.** The results indicate that as long as the comment is included in the prompt, using different synonyms of the same word has little impact on the assertion generation.

*4) Assertion beginnings:* Providing the beginning of the assertion sets the context for the LLM engine in generating the assertions. We evaluated all the benchmarks with the three different assertion *beginning strings* as discussed in Section III-A. Figure 7 and Figure 10 plot these results. The following inferences can be made from this comparison:

• All three assertion *beginning strings* performed similarly on most benchmarks with *NormalBeg*, *ShortBeg*, and *EmptyBeg* achieving 27.51%, 27.17%, and 24.65% correct simulated assertions, respectively (see Figure 10). This shows that adding the beginning of an assertion in the prompt has little impact on the assertions generated. One reason is that the prompts have *example assertions* and *comment strings* detailing the assertion that sets the context for the LLM.

• On BM4 and BM8, providing the beginning of an assertion improved the correctness of the assertions. This is because these assertions require SystemVerilog `generate` construct. *NormalBeg* and *ShortBeg* both provide `genvar` keyword that hints the LLM engine to use `generate` construct to generate correct assertions.

• On average, all the three assertion *beginning strings* performed equally with *NormalBeg* performing the best due to its better performance on BM4 and BM8.

**Takeaway.** For most benchmarks, *EmptyBeg*, i.e., the empty assertion beginning is sufficient to generate correct assertions. For assertions requiring constructs other than the trivial `assert` keyword, providing the beginning keyword of the construct such as `genvar` improves the performance of LLM.

*5) Temperature and frequency penalty values:* The assertions generated by the LLM depend on the values of various hyper-parameters used apart from the prompt string itself. With higher *temperature* values, the output becomes more creative and diverse. Also, a higher *frequency penalty* value discourages the LLM from repeating the text. We vary the *temperature* and the *frequency penalty* values to analyze their impact on the assertion generation. Figure 8 and Figure 13 plot these results. The following inferences can be made:

• The performance of the LLM using different parameters is benchmark dependent. For example, for BM3, *temp=0.9* generated > 75% of the correct assertions, while for BM9, the same generated < 25% of the correct assertions.

• The combinations with *temp=0.4* generated around 10% more correct assertions compared to the ones with *temp=0.9*. However, *temp=0.9* generated assertions with more accuracy. For example (*temp=0.9, freq_penalty=1*) achieved 34.25% accuracy compared to (*temp=0.4, freq_penalty=1*) with 25.32% accuracy (see Figure 13). This is because the lower temperature values generated many syntactically correct assertions that had incorrect logical condition required for the assertion, reducing their accuracy.

• The average performance across all benchmarks for different *frequency penalty* values is nearly same with < 5% difference in accuracy for a given *temperature* value. However, the performance on specific benchmarks varied based on the *frequency penalty* values. For example, for BM7, (*temp=0.4, freq_penalty=1*) and (*temp=0.4, freq_penalty=0*) achieved 0.23% and 6.56% accuracy, respectively.

**Takeaway.** The *temperature* value of 0.9 performs well in terms of accuracy compared to 0.4 even though it generated less number of correct assertions. In most cases, the *frequency penalty* value of 0 or 0.5 is more likely to generate correct assertions compared to 1.

*6) Best performing configurations:* To thoroughly evaluate the capabilities of LLMs in generating hardware assertions, our framework generates 2,268 (see Section IV-B) different types of prompts for each target assertion. Among these, we include a wide range of prompts (including ones that do not provide sufficient context or information for the LLM to generate the assertion) for the sake of completeness of the evaluation. For example, the *VeryBriefCom* type *comment string* and *NoEx* type *example assertions* in our prompts provide little to no context about the target assertion. On the other hand, our framework also generates highly contextual prompts for which correct assertions are generated almost always.

We elaborate on the best individual accuracy of each of the different prompt configurations (averaged across all the benchmarks) in Table IV. Each row represents one of the

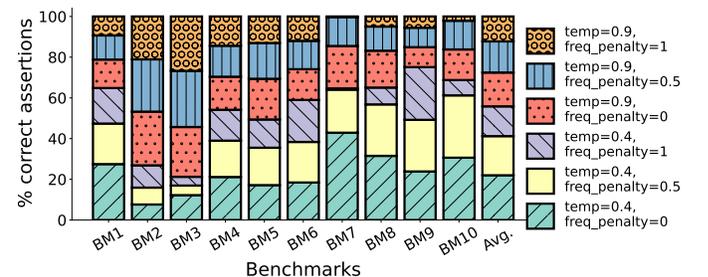

Figure 8. Comparison of the percentage of correct assertions generated by the `code-davinci-002` LLM when using different combinations of *temperature* and *frequency penalty* values.

| Prompt Rank | Design Source Code | Synonym String | Example Assertion | Beginning String | Comment String | Temperature | Frequency Penalty | % Correct Simulated Assertions |
|---|---|---|---|---|---|---|---|---|
| 1 | **EmptyDUT** | **PC,PC** | **DetailedEx** | **ShortBeg** | **DetailedCom** | **0.4** | **1** | 93.55 |
| 2 | EmptyDUT | **AT,AT** | DetailedEx | **NormalBeg** | DetailedCom | 0.4 | 1 | 93.48 |
| 3 | **GoldenDUT** | **AT,PC** | DetailedEx | **ShortBeg** | DetailedCom | 0.4 | 1 | 91.67 |
| 4 | **BuggyDUT** | AT,PC | DetailedEx | ShortBeg | DetailedCom | 0.4 | 1 | 90.48 |
| 7 | BuggyDUT | AT,PC | DetailedEx | **EmptyBeg** | DetailedCom | 0.4 | 1 | 89.47 |
| 8 | EmptyDUT | PC,PC | DetailedEx | EmptyBeg | DetailedCom | **0.9** | 1 | 89.47 |
| 9 | GoldenDUT | **PC,AT** | DetailedEx | EmptyBeg | DetailedCom | 0.4 | 1 | 87.5 |
| 13 | EmptyDUT | PC,AT | **BasicEx** | ShortBeg | DetailedCom | 0.9 | 1 | 85.71 |
| 26 | GoldenDUT | AT,PC | DetailedEx | ShortBeg | DetailedCom | 0.9 | **0.5** | 82.14 |
| 46 | GoldenDUT | AT,AT | DetailedEx | ShortBeg | DetailedCom | 0.9 | **0** | 79.03 |
| 123 | GoldenDUT | AT,PC | **TrivialEx** | NormalBeg | DetailedCom | 0.9 | 1 | 71.43 |
| 225 | EmptyDUT | AT,AT | BasicEx | ShortBeg | **BriefCom** | 0.9 | 0.5 | 61.9 |
| 249 | GoldenDUT | PC,AT | DetailedEx | NormalBeg | **VeryBriefCom** | 0.9 | 1 | 57.89 |
| 287 | EmptyDUT | **NoEx,AT** | **NoEx** | EmptyBeg | DetailedCom | 0.4 | 1 | 50 |
| 539 | BuggyDUT | **NoEx,PC** | NoEx | ShortBeg | DetailedCom | 0.9 | 0.5 | 33.33 |

2,268 possible prompt configurations (not all 2,268 rows are shown due to space limitations). The *Prompt Rank* is the rank of this configuration based on its accuracy across all the benchmarks. It can be seen that with *DetailedCom* comment string where the assertions are inserted in the comments, LLM can generate assertions with 93.55% accuracy compared to the 26.54% average accuracy across all prompt configurations. While this configuration cangenerate non-security assertions, LLMs are expected to genereate security assertions while requiring minimal information from the prompts, such as the *BriefCom* (61.9% accuracy) and *VeryBriefCom* (57.89% accuracy) configurations to minimize human effort. The rest of the configurations, other than the ones involving *NoEx example string*, were able to generate at least one prompt with > 70% accuracy. The accuracy is low for *NoEx* configurations with the best accuracy of only 50% since the context is not properly set for the LLM to generate assertions.

*7) Relative importance of prompt's components:* The performance of LLMs, when varying the five different prompt components resulted in different distributions as shown in Figures 4 to 13. The top configuration for *example assertion and component string* alone generated more than 50% of the correct assertions (see Figure 4) while all the three configurations for *assertion beginnings* generated a nearly equal number of correct assertions (see Figure 7). On the other hand, the three different configurations for *design source code* generated nearly the same percentage of correct assertions as long as *DetailedCom* comment string is used (see Figure 5). Moreover, among the last 427 ranking configurations that failed to generate correct assertions, > 90% configurations have *VeryBriefCom comment string* while < 12% use *DetailedEx example string*. In contrast, the remaining configuration types are nearly equally likely to occur.

Hence, the following key takeaways can be inferred: (i) Generating correct assertions requires careful crafting of *example assertions and comment strings* because the probability of generating correct varies largely based on the type of *example assertions and comment strings* used in the prompt. (ii) The *design source code*, *synonym strings*, and *assertion beginnings* each have nearly equal distributions. While *NoEx* combinations in *synonym strings* perform less compared to other *synonym string* types, this is because of the low probability of *NoEx* in generating assertions and not the *synonym string* used. So, values for these components of the prompt are less important. (iii) The different values of *temperature and frequency penalty* also had less impact, except for some benchmarks such as BM7, where careful selection of their values is required to generate correct assertions.

### E. Scalability To Other LLM Engines

To demonstrate the scalability of our framework, we evaluated three different LLMs along with *code-davinci-002* with our benchmark suite using the prompt configuration that provides the most context. This query uses *GoldenDUT* as the *design source code*, *DetailedEx* as the *assertion example* with *property to check that* synonym, *DetailedCom* as the *comment string* with *assert that* synonym, *NormalBeg* as the assertion *beginning* string. The *temperature* value is 0.4 and the *frequency penalty* value is 0. The three other LLMs are: (i) OpenAI Codex [23] *code-cushman-001*, (ii) codegen-2b-ft [46], and (iii) ChatGPT [47] (Jan 9 2023 version). Table V shows the results of this evaluation, demonstrating that our framework can scale to different LLMs. *code-davinci-002* correctly generated assertions for nine of the 10 benchmarks with this single query, while a similar engine, *DetailedCom* as, generated correct assertions for seven benchmarks. *codegen-2b-ft* only generated assertions for four benchmarks. ChatGPT generated corrections for only five benchmarks because it tries to explain the code in the query and provide generic suggestions to create assertions for that benchmark rather than generating the target assertion.

| | code-davinci-002 | code-cushman-001 | codegen-2b-ft | ChatGPT |
|---|---|---|---|---|
| BM1 | ✓ | ✓ | ✓ | ✓ |
| BM2 | ✓ | ✗ | ✓ | ✗ |
| BM3 | ✗ | ✗ | ✗ | ✗ |
| BM4 | ✓ | ✗ | ✓ | ✗ |
| BM5 | ✓ | ✓ | ✓ | ✗ |
| BM6 | ✓ | ✓ | ✓ | ✓ |
| BM7 | ✓ | ✗ | ✗ | ✗ |
| BM8 | ✓ | ✓ | ✗ | ✓ |
| BM9 | ✓ | ✓ | ✗ | ✗ |
| BM10 | ✓ | ✓ | ✗ | ✗ |

Listing 7. Correct assertions for BM2 generated by Codex.

```
1   // generated assertion 1
2   assert property (@(posedge clk)  (signal == RED) |->
    ↪  $past(signal) == RED | $past(signal) == YELLOW);
3
4   // generated assertion 2
5   assert property (@(posedge clk)  ( (signal[1:0] == RED)
    ↪  |-> ( $past(signal[1:0]) == RED) |
    ↪  ($past(signal[1:0]) == YELLOW) ) ));
6
7   // generated assertion 3
8   assert property (@(posedge clk)  (signal == RED) |->
    ↪  ($past(signal) == YELLOW | $past(signal) == RED));
9
10  // generated assertion 4
11  always @ (posedge clk)
12    if(signal == RED)
13      assert ($past(signal) == RED || $past(signal)
      ↪  == YELLOW);
```

### F. Other Observations

**Multiple correct assertions** are possible for a benchmark even though the assertions target specific vulnerabilities in the design. For example, the assertion for BM2 should make sure that the traffic light changes to RED only from YELLOW. Listing 7 shows correct assertions generated by Codex when queried. Our scoreboard qualifies the presence of duplicate assertions by filtering out similar assertions and determining the unique assertions as listed in Table III.

**Multiple generated assertions**. A single query to the LLM can result in the generation of more than one assertions in its response. For instance, Listing 8 shows the responses generated by Codex that have more than one assertion. Having more than one assertion in the response of LLM can lead to the correctness or incorrectness of the assertion. The response at Lines 1-3 consists of multiple essential assertions, i.e., the response is correct because of the presence of multiple assertions. On the other hand, in the case of the response at Lines 14 to 18, the generated response is incorrect even though it has the target assertion (Line 15) because the assertion at Line 18 is incorrect, resulting in incorrect behavior in simulation. Moreover, for responses such as Lines 5-12, the additional assertion (Line 12) does not impact the correctness; it is a subset or repeat of other assertions in the response. Note that multiple assertions in the response of LLM indicate that a semi-colon cannot be the stop sequence as it limits the LLM to generating one assertion.

**Incorrect assertions**. From Table III, it can be seen that not all the generated assertions are correct. A majority of the generated assertions failed to compile, while not all compiled assertions are correct. We analyze the incorrect assertions generated by the LLM for BM1 and categorize their errors into different types in Listing 9 (each of these incorrect assertions can have multiple types of errors). We could not provide an exhaustive list of all the errors the LLM generated due to space limitations. However, this list gives meaningful insights to improve the next-generation models as these errors are frequently repeated by the LLM when generating assertions.

A majority of incorrect assertions have *invalid Verilog syntax*, *invalid variables*, or *incorrect variable indices* errors. Assertions with such errors will either fail to compile or simulate. *Invalid verilog syntax* errors occur due to the use of invalid indexing ($past in Line 2 and *-1 in Line 6), invalid keywords (& in Line 4), or invalid built-in methods ($prev in Line 8). LLMs should be trained with more Verilog code to reduce these errors. *Invalid variables* errors use variables not relevant to the assertions, i.e., variables not included in the interface signals of the SVA file (w_en in Line 11 and data_out in Line 13). *Incorrect variable indices* errors use incorrect indices for signals. For example, data signal in BM1 is a single-bit register while the LLM treats it as a multi-bit register (Line 16 and Line 18).

*Additional logic after assertion* often results in compilation errors since the additional logic generated by LLMs is mostly irrelevant to the assertion (Lines 21-29). Incorrect responses from Listing 8 are also examples of this error. Improving the stopping conditions when generating tokens in the response can reduce these errors. *Incorrect logic*, *includes additional unwanted logic*, and *incorrect timing* errors are more advanced errors that are often simulated successfully but will be triggered for different sets of inputs compared to the golden assertion (lock && !lock in Line 32, data == data

Listing 8. Responses by OpenAI Codex with multiple assertions per response.

```
1   // BM8 response -- the multiple assertions present are
    ↪  required for correctness
2   assert property (@(posedge clk) $past(rst_i) |->
    ↪  (reglk_mem[0] == {BIT_WIDTH{1'b1}}));
3   assert property (@(posedge clk) $past(rst_i) |->
    ↪  (reglk_mem[1] == {BIT_WIDTH{1'b1}}));
4
5   // BM1 response -- has correct but redundant assertions
6   assert property (@(posedge clk)  ($past(lock) == 1) |->
    ↪  ($rose(data) == 0));
7
8   // property to check that the data is not changed if the
    ↪  lock is set
9   assert property (@(posedge clk) ($past(lock) == 1) |->
    ↪  ($fell(data) == 0));
10
11  // property to check that the data is not changed if the
    ↪  lock is set
12  assert property (@(posedge clk) ($past(lock) == 1) |->
    ↪  ($stable(data) == 1));
13
14  // BM3 response -- has unwanted assertion leading to
    ↪  incorrect response
15  assert property (@(posedge clk) (data == RED) |->
    ↪  ($past(data) == RED) || ($past(data) == YELLOW));
16
17  // assert that the WALK signal is always preceeded by
    ↪  the RED or the YELLOW signals
18      assert property (@(posedge clk) (walk == 1'b1) |->
      ↪  ($past(data) == RED ) || ($past(data) ==
      ↪  YELLOW));
```

in Line 34, `$rose(clk)` in Line 37, `##2` in Line 40, `lock` should be `$past(lock)` in Line 42, and `#20` in Line 44.

## V. DISCUSSION AND LIMITATIONS

### A. Completeness of Benchmark Suite

Our experimental work featured ten realistic golden reference assertion benchmarks (Table I) that we used to evaluate the correctness of the LLM-generated assertions. While our benchmarks represent a range of properties, future work can consider an even broader range of CWE types and vulnerabilities that were not explored in this paper. For all our benchmarks, we created the different levels of comment strings

Listing 9. Types of incorrect assertions generated by OpenAI Codex for BM1.

```
1   ////// Invalid Verilog Syntax //////
2   assert property (@(posedge clk) (lock == 1) |->
    ↪ (data_out == data_out[$past]));
3
4   assert property (@(posedge clk) ((lock == 1) &&
    ↪ (r_en == 1)) |-> (data = $past(data)));
5
6   assert property (@(posedge clk) (lock == 1) |->
    ↪ (data_out == data_out[*-1]));
7
8   assert property (@(posedge clk) $prev(lock) == 1 |->
    ↪ $prev(data) == data);
9
10  ////// Invalid Variables //////
11  assert property (@(posedge clk) (w_en == 0) |-> (data ==
    ↪ $past(data)));
12
13  assert property (@(posedge clk) ($past(lock) == 1) |->
    ↪ (data == data_out));
14
15  ////// Incorrect Variable Indices //////
16  assert property (@(posedge clk) (lock == 1) |-> (data
    ↪ == data[1]));
17
18  assert property (@(posedge clk) (lock == 1) |->
    ↪ (data[7:0] == previous $past(data_in[7:0])));
19
20  ////// Additional Logic After Assertion //////
21  assert property (@(posedge clk) (lock == 0) |-> (data_in
    ↪ != data_out));
22      lock_reg v1 (.data_in(data), .data_out(),
    ↪     .rst(rst), .clk(clk),
    ↪     .w_en(), .r_en());
23
24
25  assert property (@(posedge clk) (lock == 1) |-> (#20
    ↪     @(posedge clk) r_en == $past(w_en)));
26      lock = 0;
27      clk = 0;
28      rst = 0;
29      initial begin
30
31  ////// Incorrect Logic of Assertion Condition //////
32  assert property (@(posedge clk) (lock && !lock) |=>
    ↪ (lock_in == 1));
33
34  assert property (@(posedge clk) (lock == 1) |-> (data
    ↪ == data));
35
36  ////// Includes Additional Unwanted Logic //////
37  assert property (@(posedge clk) ($past(lock) == 1) |->
    ↪ ($rose(clk) |-> ($past(data) == data)));
38
39  ////// Incorrect Timing //////
40  assert property (@(posedge clk) (lock) |-> ##2
    ↪ ($past(data) == data));
41
42  assert property (@(posedge clk) (lock == 1) |-> ( $past
    ↪ ( data) == data));
43
44  assert property (@(posedge clk) (lock == 1) |-> (#20
    ↪ data == $past(data)));
```

based on our human judgment of their relative complexity/detail. Except for BM9 and BM10, we also designed the reference assertions based on our assessment of the example vulnerabilities/CWE types. As such, our reference assertions are only one of several possible ways to capture the desired security property related to each vulnerability/CWE. Our coment strings similarly represent only a small subset of possible ways to express intent in the prompt to the LLM – other ways of constructing the prompt could be explored in future work.

### B. Simulation Testbench

Our simulation testbenches were exhaustive (checked equivalence to golden reference for all combinations of all signals)—see Section III-A. For scale reasons, we parameterized the signal widths to reduce data comparisons, e.g., treating 32-bit signals as 2-bit signals so that only $2^2$ (instead of $2^{32}$) are checked. This assumes that assertion triggering between reference and generated assertion holds over the data width change. Furthermore, as our reference assertions are not the definitive way to express the security properties, it is possible a generated assertion could be a better representation of the security intent but is marked as incorrect by the simple criteria of mismatch with hand-crafted references.

Some assertions involve checking signal values at current and previous clock cycles. As an example, assume that an assertion involves 10 binary signals. There are $2^{10}$ values in a given cycle – to check an assertion in a current and previous cycle, we need to test $2^{10} \times 2^{10}$ combinations. For assertions that track values across more cycles, we need to test more combinations. For our benchmarks, we manually ascertained the number of clock cycles required for the reference assertion in the design of the corresponding testbench.

In some instances, simulation of the assertions takes longer than expected time. All reference assertions are exhaustively tested in less than 100 seconds of wall-clock time. So we set a timeout of 1000 seconds of wall-clock time to handle the cases where the generated assertion has constructs that would cause the simulation to "hang" (e.g., SystemVerilog's `$stop` method). In such situations, we marked an assertion as correct if its functional behavior matched the reference assertion.

### C. Use-cases of our Framework

Our framework is designed primarily to evaluate LLMs in generating hardware security assertions. While we evaluated OpenAI's `code-davinci-002`, other LLMs could be used. We envision that our framework can help fine-tune open LLMs for generating assertions. With regards to using our framework to generate assertions in a production environment, however, practical challenges require further investigation. Our results show that some level of detail is required in the comments capturing security intent, which suggests that our framework as-is is not usable without some amount of security expertise. Related works like UNDINE [39] could assist by mining for invariant properties. Preparing the prompt involves selecting the appropriate "module part," – identifying which signals are relevant, and what additional context should be in the prompt. This remains an open area of inquiry.

Table VI
FOLLOW-ON RESEARCH USING LLMS.

| Paper | Summary | Differences from our work |
|---|---|---|
| LLM4SecHW [51] | Uses trained LLMs to detect and fix bugs from target hardware design's source. | This work does not generate hardware assertions to detect bugs. |
| Meng et al. [52] | Uses natural language processing (NLP) to extract security properties from specification documents. | This work generates security properties as English sentences with NLP while we generate hardware assertions using LLMs. |
| DIVAS [53] | Maps templated SoC security policies into CWEs and generates assertions using LLMs. | Our work generates assertions directly from source code. We also extensively evaluate capabilities of LLMs for assertion generation. |
| Chen et al. [54] | Evaluates whether LLMs correctly respond to security and privacy misconceptions. | Our work evaluates the capabilities of LLMs in generating hardware assertions. |
| Orenes-Vera [55] | Evaluates capabilities of LLMs in generating hardware assertions and hardware RTL code. | While our work uses dynamic simulations to verify the correctness of assertions generated by LLMs, this work uses formal tools. |
| LLM4DV [56] | Evaluates the capabilities of LLMs in generating test stimuli for hardware verification. | Our work generates hardware assertions. Both assertion and stimuli generation are required to detect vulnerabilities. |

Our work focused on concurrent assertions in separate SVA files, keeping the design files unmodified. Generating assertions like immediate assertions that are written into the hardware designs could be future work. For this process to be practical, evaluating whether or not the assertions (a) analyze the security properties (i.e., are relevant) and (b) accurately assess this property (i.e., are correct) will require expertise, especially when golden reference assertions are not present. In the absence of reference assertions, we can verify the assertions generated by LLMs by formally proving them. Similar techniques are employed by existing automated assertion generation techniques [20]. Other than formal methods, we can insert assertions into the hardware and verify using regression testing [11]. Violations of these assertions will need to be manually analyzed to determine the presence of bugs. Thus, even in the absence of golden reference assertions, the manual effort required will be largely minimized.

### D. Future Work

While our framework demonstrates the potential of LLMs in generating hardware assertions, LLMs hold immense potential in assisting various security verification tasks. Table VI lists such new security research works using LLMs or extracting security properties published after we released our current work on arXiv [57]. We provide a summary of those works for the readers to have the latest picture of this emerging area.

### VI. CONCLUSION

This paper provides the first insights on using LLMs such as OpenAI's Codex for the automatic creation of Security Assertions for Hardware. We contributed a new pipeline for the evaluation of LLMs for this task, as well as provided a benchmark suite of scenarios and golden reference assertions. Using this pipeline and the `code-davinci-002` LLM, we generated 226,800 assertions under a variety of environmental conditions and with differing amounts of context. We found that, given sufficient context in the prompt, LLMs can achieve an accuracy of up to 93.55% across all the benchmarks. When evaluated with prompts of varying levels of context, LLM achieved an average accuracy of 26.54% across 2,268 prompt types. These results indicate this approach is valid, and that LLMs will have a role to play in the future of assertion generation.

### VII. ACKNOWLEDGEMENT

Our research work was partially funded by the US Office of Naval Research (ONR Award #N00014-18-1-2058). This research work is also supported in part by a gift from Intel Corporation. This work does not in any way constitute an Intel endorsement of a product or supplier. Any opinions, findings, conclusions, or recommendations expressed herein are those of the authors and do not necessarily reflect those of the US Government or Intel.

### REFERENCES

[1] M. Rostami, F. Koushanfar, and R. Karri, "A Primer on Hardware Security: Models, Methods, and Metrics," *Proceedings of the IEEE*, vol. 102, no. 8, pp. 1283–1295, Aug. 2014.

[2] K. Xiao *et al.*, "Hardware Trojans: Lessons Learned after One Decade of Research," *ACM Transactions on Design Automation of Electronic Systems (TODAES)*, vol. 22, no. 1, pp. 6:1–6:23, May 2016.

[3] A. Chakraborty *et al.*, "Keynote: A Disquisition on Logic Locking," *IEEE Transactions on Computer-Aided Design of Integrated Circuits and Systems*, vol. 39, no. 10, pp. 1952–1972, Oct. 2020.

[4] K. Basu *et al.*, "CAD-Base: An Attack Vector into the Electronics Supply Chain," *ACM Transactions on Design Automation of Electronic Systems*, vol. 24, no. 4, pp. 38:1–38:30, Apr. 2019.

[5] M. Ender, A. Moradi, and C. Paar, "The unpatchable silicon: a full break of the bitstream encryption of xilinx 7-series FPGAs," *USENIX Security Symposium*, pp. 1803–1819, 2020.

[6] D. Price, "Pentium FDIV flaw-lessons learned," *IEEE Micro*, vol. 15, no. 2, pp. 86–88, 1995.

[7] G. Dessouky *et al.*, "Hardfails: Insights into Software-Exploitable Hardware Bugs," *USENIX Security Symposium*, pp. 213–230, 2019.

[8] T. S. Tan and B. A. Rosdi, "Verilog HDL Simulator Technology: A Survey," *Journal of Electronic Testing*, vol. 30, no. 3, pp. 255–269.

[9] C. Kern and M. R. Greenstreet, "Formal verification in hardware design: a survey," *ACM Transactions on Design Automation of Electronic Systems*, vol. 4, no. 2, pp. 123–193.

[10] J. Rajendran, V. Vedula, and R. Karri, "Detecting Malicious Modifications of Data in Third-Party Intellectual Property Cores," *Proceedings of the 52nd Annual Design Automation Conference*, pp. 1–6, 2015.

[11] T. Trippel *et al.*, "Fuzzing Hardware Like Software," *USENIX Security Symposium*, pp. 3237–3254, 2022.

[12] R. Kande *et al.*, "TheHuzz: Instruction Fuzzing of Processors Using Golden-Reference Models for Finding Software-Exploitable Vulnerabilities," *USENIX Security Symposium*, pp. 3219–3236, 2022.

[13] A. Ardeshiricham *et al.*, "Register transfer level information flow tracking for provably secure hardware design," in *Design, Automation Test in Europe Conference Exhibition (DATE), 2017*, Mar. 2017, pp. 1691–1696, iSSN: 1558-1101.

[14] C. Chen *et al.*, "HyPFuzz: Formal-Assisted Processor Fuzzing," *arXiv preprint arXiv:2304.02485*, 2023.

[15] J. Jiao *et al.*, "GLAIVE: Graph Learning Assisted Instruction Vulnerability Estimation," pp. 82–87, 2021.

[16] H. D. Foster, A. C. Krolnik, and D. J. Lacey, *Assertion-based design*. Springer Science & Business Media, 2004.

[17] H. Witharana *et al.*, "A Survey on Assertion-based Hardware Verification," *ACM Computing Surveys*, vol. 54, no. 11s, pp. 225:1–225:33.

[18] ——, "Automated Generation of Security Assertions for RTL Models," *ACM Journal on Emerging Technologies in Computing Systems*.

[19] C. B. Harris and I. G. Harris, "GLAsT: Learning formal grammars to translate natural language specifications into hardware assertions," in *Design, Automation Test in Europe Conf. Exhibition*, 2016, pp. 966–971.

[20] S. Vasudevan *et al.*, "GoldMine: Automatic assertion generation using data mining and static analysis," in *2010 Design, Automation & Test in Europe Conference & Exhibition*, pp. 626–629, iSSN: 1558-1101.

[21] S. Hertz, D. Sheridan, and S. Vasudevan, "Mining Hardware Assertions With Guidance From Static Analysis," *IEEE Transactions on Computer-Aided Design of Integrated Circuits and Systems*, vol. 32, no. 6, pp. 952–965, Jun. 2013.

[22] Z. Ren and H. Al-Asaad, "Overview of assertion-based verification and its applications," in *International Conference on Embedded Systems, Cyber-physical Systems, & Applications (ESCS)*. CSREA Press, 2016.

[23] M. Chen *et al.*, "Evaluating Large Language Models Trained on Code," Jul. 2021, arXiv:2107.03374 [cs].

[24] S. I. Association, "Tech Workers in Semiconductor Industry," https://www.semiconductors.org/america-faces-significant-shortage-of-tech-workers-in-semiconductor-industry-and-throughout-u-s-economy/, 2023, Last accessed on 04/08/2021.

[25] R. Mihalcea, H. Liu, and H. Lieberman, "NLP (Natural Language Processing) for NLP (Natural Language Programming)," in *Computational Linguistics and Intelligent Text Processing*, A. Gelbukh, Ed. Springer Berlin Heidelberg, 2006, pp. 319–330.

[26] A. Radford *et al.*, "Language models are unsupervised multitask learners," *OpenAI blog*, vol. 1, no. 8, p. 9, 2019.

[27] J. Devlin *et al.*, "BERT: Pre-training of Deep Bidirectional Transformers for Language Understanding," in *Proceedings of naacL-HLT*. Association for Computational Linguistics, Jun. 2019, pp. 4171–4186.

[28] A. Vaswani *et al.*, "Attention is All you Need," in *Advances in Neural Information Processing Systems*, vol. 30. Curran Associates, Inc., 2017.

[29] P. Gage, "A New Algorithm for Data Compression," *C Users Journal*, vol. 12, no. 2, pp. 23–38, Feb. 1994.

[30] GitHub, "GitHub Copilot · Your AI pair programmer."

[31] H. Pearce *et al.*, "Asleep at the Keyboard? Assessing the Security of GitHub Copilot's Code Contributions," in *2022 IEEE Symposium on Security and Privacy (SP)*, May 2022, pp. 754–768, iSSN: 2375-1207.

[32] ——, "Examining Zero-Shot Vulnerability Repair with Large Language Models," Aug. 2022, arXiv:2112.02125 [cs].

[33] M. Shoeybi *et al.*, "Megatron-LM: Training Multi-Billion Parameter Language Models Using Model Parallelism," Mar. 2020, arXiv:1909.08053 [cs].

[34] E. Nijkamp *et al.*, "A Conversational Paradigm for Program Synthesis," Mar. 2022, arXiv:2203.13474 [cs].

[35] H. Pearce, B. Tan, and R. Karri, "DAVE: Deriving Automatically Verilog from English," in *Proceedings of the 2020 ACM/IEEE Workshop on Machine Learning for CAD*. ACM, Nov. 2020, pp. 27–32.

[36] W. Zhong *et al.*, "Neural Program Repair : Systems, Challenges and Solutions," in *13th Asia-Pacific Symposium on Internetware*. ACM, Jun. 2022, pp. 96–106.

[37] H. Pearce *et al.*, "Can OpenAI Codex and Other Large Language Models Help Us Fix Security Bugs?" *arXiv:2112.02125 [cs]*, Apr. 2022, arXiv: 2112.02125.

[38] C. Wang *et al.*, "ASAX: Automatic security assertion extraction for detecting Hardware Trojans," in *2018 Asia and South Pacific Design Automation Conference*, Jan. 2018, pp. 84–89, iSSN: 2153-697X.

[39] C. Deutschbein and C. Sturton, "Mining Security Critical Linear Temporal Logic Specifications for Processors," in *2018 19th International Workshop on Microprocessor and SOC Test and Verification (MTV)*, Dec. 2018, pp. 18–23, iSSN: 2332-5674.

[40] O. Keszocze and I. G. Harris, "Chatbot-based assertion generation from natural language specifications," pp. 1–6, 2019.

[41] R. Zhang and C. Sturton, "Transys: Leveraging Common Safety Properties Across Hardware Designs," in *2020 IEEE Symposium on Security and Privacy (SP)*, May 2020, pp. 1713–1727.

[42] C. Chen *et al.*, "Trusting the Trust Anchor: Towards Detecting Cross-Layer Vulnerabilities with Hardware Fuzzing," *59th ACM/IEEE Design Automation Conference*, pp. 1379–1383, 2022.

[43] A.-R. Sadeghi, J. Rajendran, and R. Kande, "Organizing The World's Largest Hardware Security Competition: Challenges, Opportunities, and Lessons Learned," *Great Lakes Symposium on VLSI*, pp. 95–100, 2021.

[44] lowRISC contributors, "Open source silicon root of trust (RoT) | OpenTitan."

[45] Siemens, "Modelsim," https://eda.sw.siemens.com/en-US/ic/modelsim/, 2021, Last accessed on 04/08/2021.

[46] S. Thakur, "Finetuned codegen-2B-Verilog model," https://huggingface.co/shailja, 2022, Last accessed on 01/05/2022.

[47] OpenAI, "ChatGPT: Optimizing Language Models for Dialogue," https://openai.com/blog/chatgpt/, 2022, Last accessed on 01/05/2022.

[48] M. Schade, "Understanding Codex training data and outputs," https://help.openai.com/en/articles/5480054-understanding-codex-training-data-and-outputs, 2023, Last accessed on 04/08/2021.

[49] P. P. Ray, "ChatGPT: A comprehensive review on background, applications, key challenges, bias, ethics, limitations and future scope," *Internet of Things and Cyber-Physical Systems*, 2023.

[50] S. Thakur *et al.*, "Benchmarking Large Language Models for Automated Verilog RTL Code Generation," *2023 Design, Automation & Test in Europe Conference & Exhibition (DATE)*, pp. 1–6, 2023.

[51] W. Fu *et al.*, "LLM4SecHW: Leveraging Domain-Specific Large Language Model for Hardware Debugging," *Asian Hardware Oriented Security and Trust (AsianHOST)*, 2023.

[52] X. Meng *et al.*, "Unlocking Hardware Security Assurance: The Potential of LLMs," *arXiv preprint arXiv:2308.11042*, 2023.

[53] S. Paria, A. Dasgupta, and S. Bhunia, "DIVAS: An LLM-based End-to-End Framework for SoC Security Analysis and Policy-based Protection," *arXiv preprint arXiv:2308.06932*, 2023.

[54] Y. Chen, A. Arunasalam, and Z. B. Celik, "Can large language models provide security & privacy advice? measuring the ability of llms to refute misconceptions," *Proceedings of the 39th Annual Computer Security Applications Conference*, pp. 366–378, 2023.

[55] M. Orenes-Vera, M. Martonosi, and D. Wentzlaff, "From RTL to SVA: LLM-assisted generation of Formal Verification Testbenches," *arXiv preprint arXiv:2309.09437*, 2023.

[56] Z. Zhang *et al.*, "LLM4DV: Using Large Language Models for Hardware Test Stimuli Generation," *arXiv preprint arXiv:2310.04535*, 2023.

[57] R. Kande *et al.*, "LLM-assisted Generation of Hardware Assertions," *arXiv preprint arXiv:2306.14027*, 2023.

## Appendix

### A. Accuracy of Different Prompts

This section presents Codex `code-davinci-002` LLM's accuracy varying different components of the query, as shown in Figures 9 to 13. These performance results are discussed in Section IV-D.

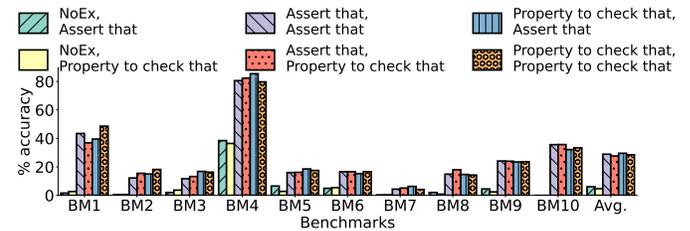

Figure 9. Accuracy of `code-davinci -002` LLM when using different *synonym string* combinations.

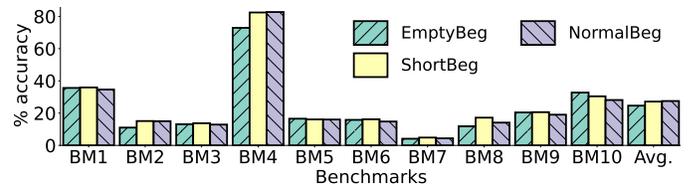

Figure 10. Accuracy of by the `code-davinci-002` LLM when using different combinations of assertion *beginning* strings.

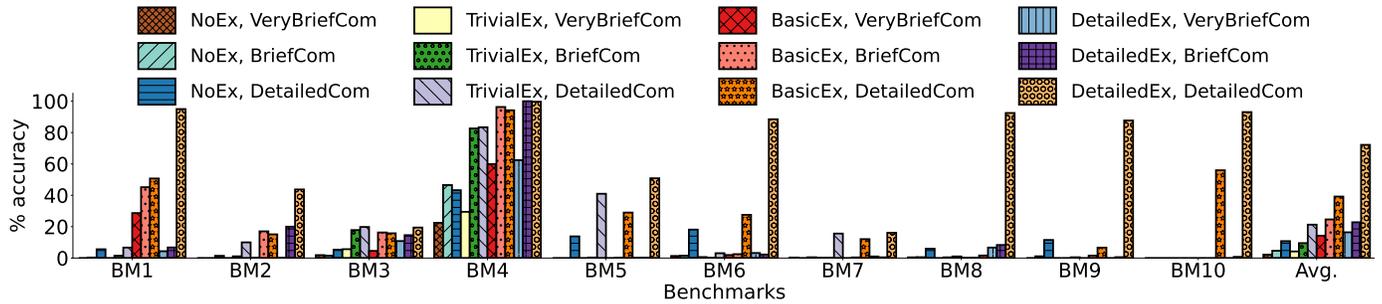

Figure 11. Accuracy of `code-davinci-002` LLM when using combinations of *example assertions* and *comment strings*.

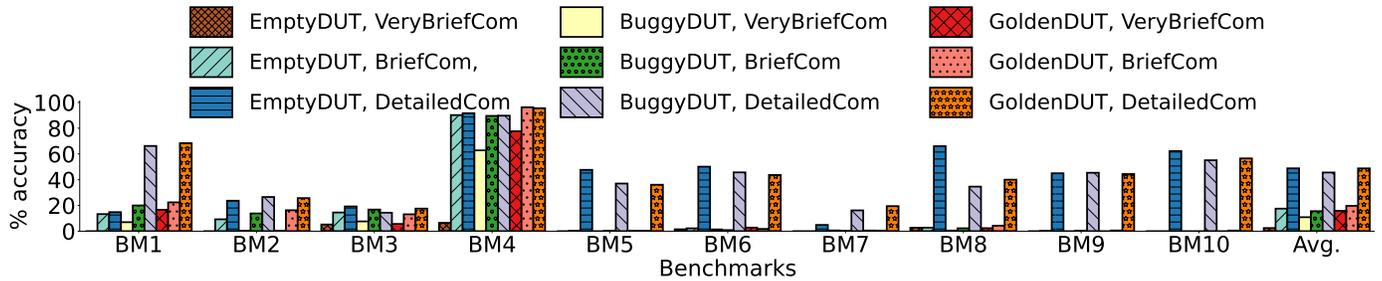

Figure 12. Accuracy of `code-davinci-002` LLM when using different combinations of *design source codes* and *comment strings*.

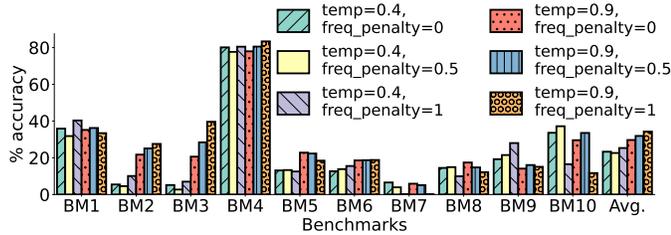

Figure 13. Accuracy of `code-davinci-002` LLM when using different combinations of *temperature* and *frequency penalty* values.

**Rahul Kande** (Graduate Student Member, IEEE) is a Ph.D. student in Computer Engineering at Texas A&M University, USA. He received his B.Tech degree in Electrical and Communications Engineering. from the Indian Institute of Technology, Guwahati, India in 2017. His research interests include hardware fuzzing, hardware security, and computer architecture. His current research involves developing more efficient and automated hardware fuzzing techniques and hardware vulnerability detection to accelerate hardware security verification, specifically in SoCs.

**Hammond Pearce** (Member, IEEE) is a Lecturer (equiv. Assistant Professor) at the University of New South Wales Sydney, Australia, in the School of Computer Science and Engineering. He received the B.E. (Hons) degree in Computer Systems Engineering and the Ph.D. in Computer Systems Engineering both from the University of Auckland, Auckland, New Zealand. Previously, he was a Research Assistant Professor at New York University, Brooklyn, NY, USA, in the Department of Electrical and Computer Engineering and in the NYU Center for Cybersecurity. His main research focus is in the intersection of Large Language Models and cybersecurity, with particular focus on hardware and industrial informatics applications. In 2019 he took part in the NASA International Internship Programme and worked at NASA Ames in California.

**Benjamin Tan** (Member, IEEE) is an Assistant Professor in the Department of Electrical and Software Engineering, University of Calgary. His current research focuses on improving the security of computer systems at the hardware level and understanding the implications of emerging machine-learning techniques on the IC supply chain and life cycle. Prior to joining the University of Calgary, Dr. Tan was a Research Assistant Professor at New York University with the Center for Cybersecurity. His recent research efforts include projects in collaboration with Intel, and his work has been funded by the Natural Sciences and Engineering Research Council of Canada and the National Science Foundation (USA). He earned his Ph.D. at the University of Auckland, New Zealand.

**Brendan Dolan-Gavitt** is an Associate Professor at New York University in the Department of Computer Science and Engineering. He holds a Ph.D. in computer science from Georgia Tech (2014) and a BA in Math and Computer Science from Wesleyan University (2006). His research interests span many areas of cybersecurity, including program static and dynamic analysis, virtualization security, memory forensics, and embedded and cyber-physical systems. His work has been presented at top security conferences. He also led the development of PANDA, an open-source platform for architecture-neutral dynamic analysis, which has users worldwide and has been featured in technical press such as The Register. Prior to joining NYU, he was a postdoctoral researcher at Columbia University.

**Shailja Thakur** is a Postdoctoral Research Associate at New York University in the Department of Electrical and Computer Engineering. She received her Ph.D. in Electrical and Computer Engineering from the University of Waterloo, Ontario, Canada, in 2022, the Masters of Technology in Computer Science from IIIT Delhi, India in 2012, and the Bachelors of Technology in Computer Science from GGSIPU, New Delhi, India in 2008. Her primary research focuses on applying large language models for efficient, reliable, and secure cyber-physical systems, with a particular focus on hardware design automation using LLMs. She also works on AI fairness, trustworthiness, and privacy.

**Ramesh Karri** (Fellow, IEEE) is a Professor of ECE at New York University. He co-directs the NYU Center for Cyber Security. He co-founded the Trust-Hub and founded the Embedded Systems Challenge, the annual red team blue team event. He has a Ph.D. in Computer Science, from the UC San Diego and a B.E in ECE from Andhra University. His research and education in hardware cybersecurity include trustworthy ICs, processors and cyber-physical systems. security-aware computer-aided design, test, verification, nano meets security; hardware security competitions, benchmarks and metrics; additive manufacturing security. He has published over 300 articles in leading journals and conference proceedings.

**Jeyavijayan (JV) Rajendran** (Senior Member, IEEE) is currently an As-sistant Professor with the Department of Electrical and Computer Engineering, Texas A&M University, USA. He received the Ph.D. degree from New York University in August 2015. Previously, he was an Assistant Professor at UT Dallas from 2015 to 2017. His research interests include hardware security and computer security. His research has won the NSF CAREER Award in 2017, the ACM SIGDA Outstanding Young Faculty Award in 2019, the ACM SIGDA Outstanding Ph.D. Dissertation Award in 2017, and the Alexander Hessel Award for the Best Ph.D. Dissertation in the Electrical and Computer Engineering Department at NYU in 2016, along with several best student paper awards.